\documentclass{elsart}

\usepackage{natbib}

\usepackage{graphicx}

\usepackage{amssymb}

\newcommand{\fsize}{0.8\textwidth}

\begin{document}

\begin{frontmatter}



\title{Optimal boarding method for airline passengers}


\author{Jason H. Steffen\thanksref{label2}}
\ead{jsteffen\_AT\_fnal.gov}
\thanks[label2]{Brinson Postdoctoral Fellow, Fermilab Center for Particle Astrophysics}

\address{MS 127, PO Box 500 \\ Batavia, IL 60510}

\begin{abstract}
Using a Markov Chain Monte Carlo optimization algorithm and a computer simulation, I find the passenger ordering which minimizes the time required to board the passengers onto an airplane.  The model that I employ assumes that the time that a passenger requires to load his or her luggage is the dominant contribution to the time needed to completely fill the aircraft.  The optimal boarding strategy may reduce the time required to board and airplane by over a factor of four and possibly more depending upon the dimensions of the aircraft.  In addition, knowledge of the optimal boarding procedure can inform decisions regarding changes to methods that are employed by a particular carrier.  I explore some of the salient features of the optimal boarding method and discuss practical modifications to the optimal.  Finally, I mention some of the benefits that could come from implementing an improved passenger boarding scheme.
\end{abstract}




\end{frontmatter}

\section{Introduction}
Several passenger boarding schemes are used by the airline industry in effort to quickly load passengers and their luggage onto an airplane.  Since the passenger boarding time often takes longer than refueling and restocking the airplane its reduction could constitute a significant savings to a particular carrier, especially for airplanes which make several trips in a day.

Conventional wisdom would suggest that boarding from the front to the back is the worst case but that boarding from the back to the front is optimal or nearly so.  Indeed, this is the strategy that is often employed, boarding passengers in blocks from the rear of the plane to the front.  In this case, conventional wisdom only provides an answer that is half right.  The worst boarding method is, indeed, to board the plane from front to back.  As I will show however, boarding the airplane from the back to the front is very likely the \textit{second worst} method.

Previous studies of the airplane boarding used a variety of approaches.  Pioneering work done by \cite{belgians} used computer simulations of several boarding schemes and found there is much room for improvement over traditional back-to-front boarding.  Of particular interest is that they point out that random boarding is superior to many traditional methods.  A later computer simulation study \citep{vandenbriel} confirmed that traditional methods are not optimal, even in light of different assumptions regarding the primary cause of delay in the boarding process.  Analytic work by \cite{bachmat}, which modeled optimal airplane boarding as an extremal path in a two-dimensional, Lorentzian geometry, broadly confirmed the findings of \cite{belgians} and was able to interpret those findings in the context of their model---thus providing an explanation for why the different boarding strategies perform as they do relative to each other.

Each of these studies, particularly the computational works, focused on different components of delay.  Consequently, while all found traditional methods lacking, none agree on the best overall approach.  A portion of their disagreement arises from differing assumptions.  Another point of discrepancy stems from the fact that their focus was largely on finding the best practical approach or the best of a set of approaches.  This limitation on the scope of previous works meant that they didn't always study the same boarding methods.

Thus, the question remains, ``what method, practical or otherwise, gives the fastest boarding time?''  It is this question that I address here.  The answer to this question, whether directly applicable or not, is valuable as it can inform decisions regarding the worth changing an existing policy since it indicates how much room there is for improvement.  Moreover, once the optimal boarding method is identified one can explore the reasons why it is successful and incorporate those characteristics into a more practical scheme.

In this work I infer from the results of a computer simulation and an associated optimization algorithm the optimal passenger boarding method.  The fundamental assumption that I make is that the bulk of the time to load the airplane is consumed by time that it takes the passengers to load their luggage.  All other effects, such as the time used by passengers who stand up to retrieve an item, who sit in the wrong seat, or who must pass by someone that is already seated in their row are initially treated as negligible.  A more sophisticated model could include such detail; or, many of their effects could be folded into the model via some of the available parameters or components (such as the distribution from which the passenger luggage loading times are selected).

With the stated assumption, I find that the boarding time for the optimal scheme can be significantly faster than the boarding time of the worst case---between a factor of 4 and 10 faster depending upon the length of the airplane and other model parameters.  In this article I describe the techniques used to find the optimal method, I interpret the results and use that interpretation to discuss the merits of some schemes that are employed by the industry, finally I give some concluding remarks.  Note that I generally use the term ``boarding'' to refer to the boarding process itself and the term ``loading'' to refer to the passengers loading their luggage.  Thus, boarding time and loading time are the times required to fill the aircraft and the time required to load one's luggage respectively.

\section{Analysis Approach}

\subsection{Airplane and Passenger Models}

The nominal airplane model that I use seats 120 passengers with six passengers per row and 20 rows.  Since the focus is on the general boarding procedures, there is no first-class cabin, no priority seating, and each flight is completely full.  I discuss the effects of changes to this airplane model in section \ref{results}.

The passengers are each assigned a seat and the number of time steps that they need to load their luggage, a random number between 0 and 100 unless otherwise stated.  Other, human nature assumptions include: 1) that a person will not move unless there is enough space between them and the person in front of them---two steps in this case, 2) that if they are moving, then they will occupy any empty space in front of them prior to stopping (this results in passengers bunching-up again as they come to a halt), 3) that they require one space either in front of or behind them in order to load their luggage (this is related to the congestion parameter $k$ introduced by \cite{bachmat}), and 4) that they only load their luggage into the bins above their assigned row.  In section \ref{results} I discuss the effects of changing any of these parameters or the distribution from which the luggage loading times are assigned.

This model does not include the effects of aisle vs. window seats, the clustering of passengers into companions or families, and other effects of human nature.  While adding these features might improve the accuracy of the results, these effects are not likely to be the primary issue and consequently should not be the fundamental concern when finding the general strategy for a passenger boarding scheme.  Moreover, many of their effects can be accounted for once the optimal boarding method, based upon the stated assumptions, is identified.  I discuss these issues in sections \ref{robust} and \ref{practical}.

\subsection{Optimization Algorithm}

The algorithm that I use to find the optimal loading order is based upon a Markov Chain Monte Carlo (MCMC) algorithm and is similar to the METROPOLIS algorithm \citep{metropolis}.  Starting with an initial passenger order I load the airplane and record the loading time.  Then, starting with that initial order, the positions of two random passengers are exchanged and the airplane is loaded again.  If the airplane boards as fast or faster than the previous iteration, then I accept the current passenger order, swap the positions of two additional random passengers, and repeat the process.  If the current configuration loads more slowly than the previous, then I reject the change, return to the previous configuration, and repeat the process beginning there.  I stop after $\sim 10,000$ iterations since adding additional steps does not significantly change the results.

Unlike a traditional MCMC, I do not allow any configuration which loads more slowly to be accepted.  Technically, this means that the stated algorithm only finds a local minimum.  However, when I include this aspect the results are unchanged while the time needed to converge increases.  Thus, the choice to neglect that aspect of an MCMC analysis should not affect the results stated here.

That being said, it is possible for many configurations to board in the same time or to be near enough that the differences in boading time are not important.  This shows that a class of configurations that are effectively equivalent is more important than a single, optimum order.  For example, there is no difference in the loading time if the two aisle-seat passengers in the same row are swapped, and there is little difference in swapping two random passengers.  To identify the class of optimal configurations I tabulate the differences in seat number between adjacent passengers.  It is this distribution of seat number differences, the ``seating distribution'', that remains effectively constant from one optimization run to the next.  This fact illustrates the important point that the actual seat numbers of two adjacent passengers is less important than how far apart they sit from each other.

\section{Results\label{results}}

\begin{figure}
\begin{center}
\includegraphics[width=\fsize]{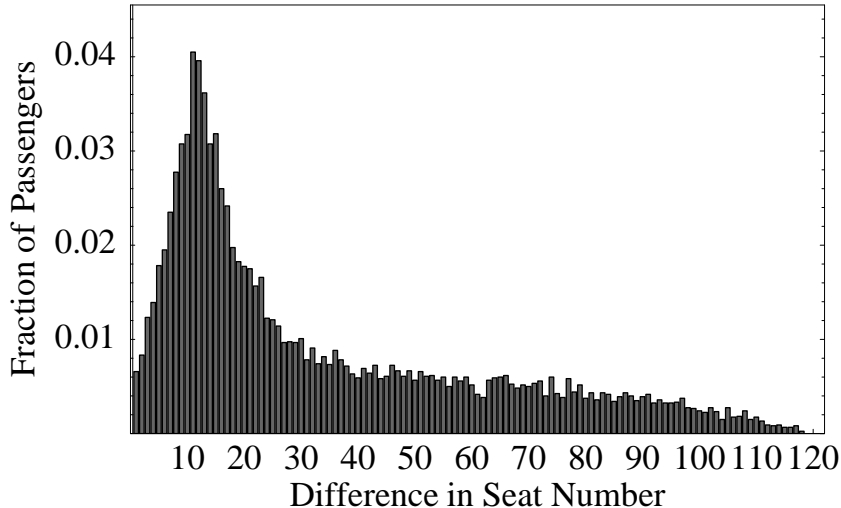}
\caption{Example of the resulting seating distribution obtained from 100 realizations of the luggage loading time distribution.}
\label{seatdist}
\end{center}
\end{figure}

The results of applying the above analysis gives the seating distribution shown in Figure \ref{seatdist}.  These results are from 100 realizations of the boarding scenario where the passengers are reassigned their luggage loading time for each of the realizations.  The largest feature is the peak near a seat difference of 12.  That difference corresponds to two rows, or the distance that I assume neighboring passengers require to load their luggage.  Other features of the peak, aside from its location, are its shape, its height relative to the rest of the distribution, and its width.  All four of these aspects depend upon the passenger and airplane model parameters and each of them could be calibrated with data as I describe below.

\begin{figure}
\begin{center}
\includegraphics[width=\fsize]{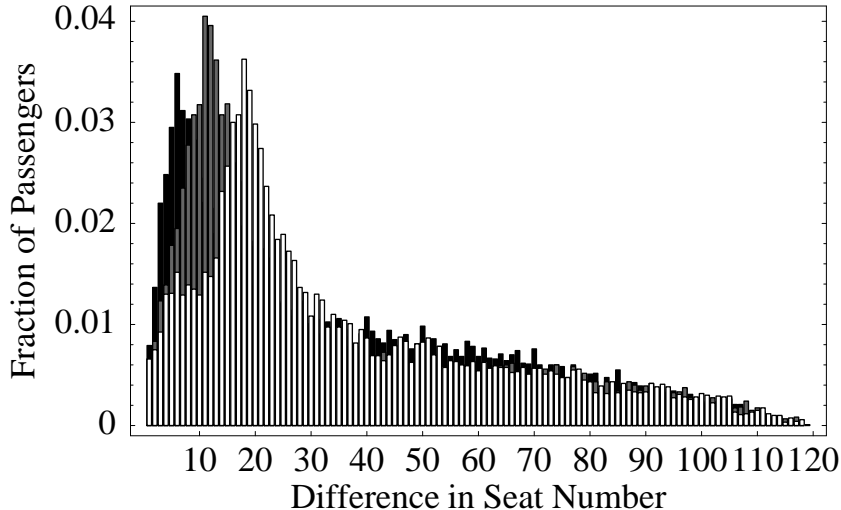}
\caption{Changes in the seating distribution as a function of the required ``personal space'' of the passengers.  This shows the distribution if no space is required (black), a single row is required (gray), and two rows are required (white).  Each distribution is calculated from 100 realizations of the luggage loading times.}
\label{personal}
\end{center}
\end{figure}

As stated, the location of the peak corresponds to the distance required by a passenger to load his luggage.  If passengers need only the space that corresponds to their assigned row, then the location of the peak would shift to a value of 6 or one row of difference.  If passengers require three rows of loading space (including their assigned row), then the peak shifts to 18.  This effect can be seen in Figure \ref{personal} where I make these changes while leaving all other model parameters fixed.
The peak of the seating distribution is symmetrical near its apex.  This is because the passengers can load their luggage using either the space in front of them or the space behind them.  The width of the peak is related to the number of seats per row.  If there are only four seats per row, then the width of the peak is more narrow.  If there are eight seats per row (still with only one aisle), then it is more broad.

The height of the peak depends upon the time that the passengers take to load their luggage.  If passengers load their luggage instantaneously, then the peak disappears altogether.  As the luggage loading time increases the penalty for having someone out of order increases and the algorithm forces more passengers to be separated by amounts nearer the minimum row separation required to load their luggage (here between 7 and 18 seats).  The height of the peak ultimately saturates when the average luggage loading time approaches the time to walk the length of the airplane.  This means that if it takes significantly longer for a typical passenger to walk the length of the airplane than it does for him to load his luggage, then the passenger ordering is much less important.

\subsection{The Optimal Loading Method}

\begin{figure}
\begin{center}
\includegraphics[height=\fsize]{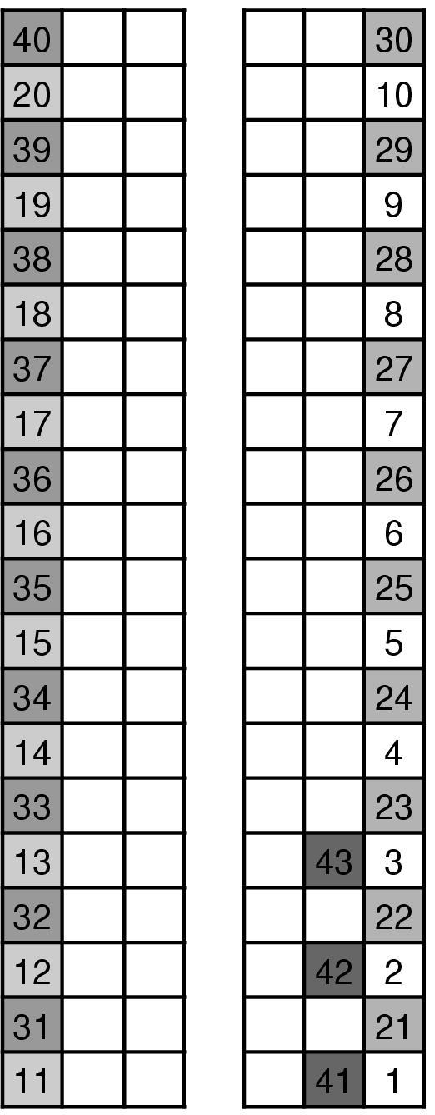}
$\qquad \qquad$
\includegraphics[height=\fsize]{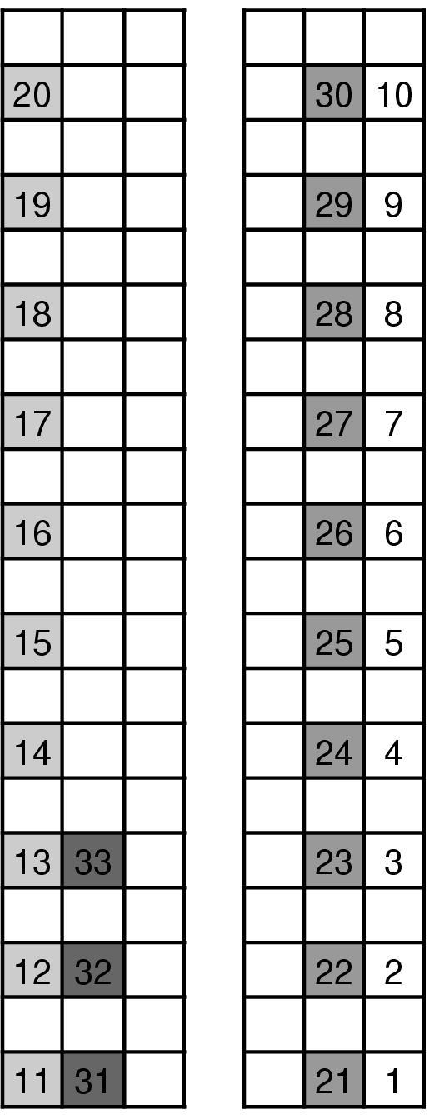}
\caption{Examples of the optimal passenger ordering---there are other permutations which would give identical results.  The seating would proceed following the patterns illustrated here.  The shading only indicates the passengers that would be inside the airplane at the same time.}
\label{seatdiagram}
\end{center}
\end{figure}

The optimal boarding scheme is found by extrapolating from the results given above, using the insight that they provide.  The reason that these seating distributions load faster than the worst case is that they allow multiple passengers to load their luggage at the same time.  The peak occurs at a distance that corresponds exactly to the space needed by adjacent passengers to do so.  Taking this to an extreme, we wish to find the configuration that allows the maximum number of passengers to load their luggage at all times during the boarding process.  That is the case where all adjacent passengers are assigned seats that are separated by exactly two rows such that the person at the front of the line is assigned a seat on the back row.  The second-order effect of windows vs. aisle seats can be incorporated here by having the passengers in the window seats enter the airplane first, then the middle seats, then the aisle seats.  The resulting seat ordering could be one of those shown in Figure \ref{seatdiagram} though there are many equivalent orderings.  This ordering scheme provides nearly a five-fold reduction in the time that it takes over the worst case---an improvement that gets larger with a larger aircraft.

A look at the optimal boarding method shows why loading from the back of the plane to the front does not provide any benefit.  If the back two rows of passengers were to board the airplane first, they would occupy roughly 12 rows of the aisle.  All but the first few would be putting their luggage away while the others waited their turn; the passengers load their luggage serially.  The optimal boarding strategy uses this aisle space more efficiently since each member of the first group of passengers who enter the airplane (10 passengers in the fiducial model) can put their luggage away; they load their luggage in parallel.  In this manner the aisle is not used as a passive extension of the waiting area, but rather as a place for passengers to actively situate themselves.  Ideally, all of the passengers that are inside the aircraft should either be seated or be loading their luggage; with none waiting.

One question that arises from this is whether or not it is practical to implement the optimal boarding scheme, where each passenger enters the airplane in a particular order.  Such a scenario may well be possible since Southwest Air has recently implemented a similar policy, at least to some extent.  Given that, however, there will always be some fraction of the passengers who are out of order; there will always be families or other groups who board together regardless of their assignments.  These same issues will affect other boarding methods in similar ways.  In the next section I test the robustness of the optimal boarding method under these circumstances.  Section \ref{practical} is a comparison of a few practical boarding methods where the passengers board in groups but are ordered randomly within those groups.

\section{Robustness of the Optimal\label{robust}}

To test the robustness of the optimal boarding scheme I conducted two experiments.  The first is to change the distribution from which I select the passenger's loading time.  The second test is to make random changes to the passenger ordering.  These changes include swapping the locations of several random pairs of passengers and shifting the entire line by some random number (moving people at the end of the line to the front).

\subsection{Changes to Loading Time Distribution}

To test the effect of a different distribution of luggage loading time, I ran my minimization software on 100 realizations of each of several distributions.  These distributions include a uniform distribution with a given mean, a normal distribution with the same mean and with a variance equal to that mean (essentially a Poisson distribution), and an exponential distribution with the same mean.  For each of these cases the resulting passenger seating distributions were statistically indistinguishable as shown by a Kolmogorov-Smirnov test \citep{press}.  Moreover, the time required to load the entire plane is not affected by these different distributions; it depends primarily upon the mean luggage loading time.

The fact that the results of this analysis does not depend upon the distribution from which the luggage loading times were chosen indicates that the optimal boarding method does not depend upon the presence of non-Gaussian outlier disturbances (as it should not).  Thus, if a particular passenger takes an unusually long time to load his luggage, the optimal method of passenger boarding is still, on average, optimal.  Moreover, since one cannot generally identify which passenger---if it is a passenger---will be the cause of delay and since that passenger's seat will be at a random location within the aircraft, the best that one can robustly expect to do is to board optimally both before and after the disturbance.

\subsection{Random Shifts and Swaps}

Randomly shifting the line does not significantly affect the boarding time.  This is because it only changes the starting point of the boarding process.  All of the passengers keep their 12 seat spacing from their neighbors and so the advantage of the optimal boarding scheme is preserved.

\begin{figure}
\begin{center}
\includegraphics[width=\fsize]{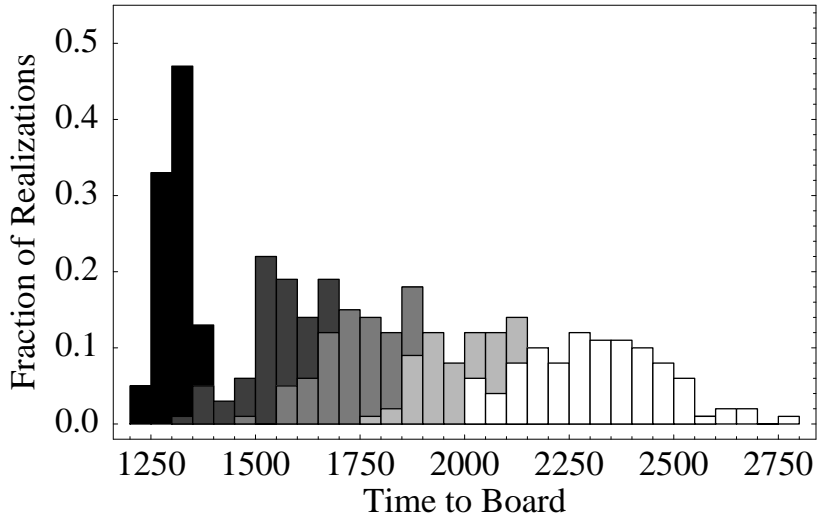}
\caption{Boarding times for 100 realizations of passengers when pairs of passengers are swapped.  The black is optimal with no swapping, then there are the histograms which correspond to 10\% (dark gray), 20\% (medium gray), 40\% (light gray), and 60\% (white) swapping.  10\% means that 6 pairs of passengers are exchanged out of 120 passengers.  The mean boarding times for these scenarios are 1312, 1585, 1795, 2084, and 2311 counts respectively.}
\label{passengermix}
\end{center}
\end{figure}

If pairs of passengers are swapped, which effectively randomizes portions of the line, then the time to board the airplane can change significantly.  Indeed, a 20\% increase in the boarding time results from randomly swapping only 10\% of the passengers---that is 6 pairs for the case of 120 passengers.  However, there is an upper limit to the adverse results of swapping passengers.  Once they are completely randomized additional swaps preserve the random nature of the passenger ordering and it doesn't get any worse.  Interestingly, random boarding takes much less than half the time of the worst case boarding; indicating that randomization is not catastrophic, a finding also identified by \cite{belgians}.  Indeed, we will see in the next section that random boarding compares favorably with traditional boarding techniques.  The effects of swapping pairs of passengers is shown in Figure \ref{passengermix}.

\section{Practical Comparison\label{practical}}

While the optimal scheme would produce the fastest boarding times, there are issues of practicality to consider.  It may be challenging to arrange all of the passengers in the proper order---though, as mentioned, at least one airline has implemented this policy.  Regardless, most airlines board the airplane in groups, presumably out of convenience and in effort to reduce confusion.  In this section I introduce a few practical modifications to the optimal boarding scheme and choose one to compare with existing methods.

\subsection{Practical Modification to the Optimal}

The advantage of the optimal boarding method comes from the fact that neighboring passengers do not sit near each other and consequently can load their luggage simultaneously.  One way to accomplish a similar effect while allowing blocks of passengers to board is to have each block contain passengers from widely separated rows.  After trying several possibilities I chose to use, as a modification to the optimal method, blocks of three consecutive seats separated by 12.

This scenario has four boarding groups and is equivalent to calling all passengers that are in even rows and from one side of the airplane.  The three remaining groups are for the other side of the airplane in the same row, then the two sides of the odd rows.  The loading time that results from this scheme is not as fast as the optimum, indeed it took about twice as long to board, but it was more than a factor of two faster than the worst case.  I call this boarding method the ``modified optimal'' method.  Similar boarding methods, arrived at by very different means, are suggested by \cite{belgians} and \cite{bachmat} and these methods also prove successful in the respective studies.

\subsection{Comparison with Other Methods}

I selected two group boarding strategies to compare with the optimal, the modified optimal, the worst case, and the second worst case (boarding from the back to the front with the passengers in order).  These are: 1) ordered blocks, where a fourth of the cabin is loaded at a time starting in the back and moving to the front and 2) a scheme where the windows are boarded first, then the middle, then the aisle seats.  Within each of these groups the travelers were randomly distributed.  Tests which used unordered blocks (blocks of 5 rows but not in order from back to front) gave similar results to the ordered blocks and were therefore not included.

The ordered block scenario reduced the boarding time to 74\% of the worst case.  Since the main contributor to fast boarding is having multiple passengers load their luggage at once, this method suffers from the fact that only passengers within a small portion of the airplane are boarding at a given time.  Within the boarding block there is the possibility of multiple people loading their luggage at once, but it is relatively small since at most three passengers (out of 30) could be simultaneously loading their luggage.

The windows-middle-aisle approach reduced the boarding time to 43\% of the worst case.  The advantage here is that passengers from anywhere in the cabin are allowed to board.  Thus, many people can load their luggage at once and the probability having two passengers from adjacent rows near each other is small.  Moreover, this probability decreases as the length of the airplane increases.  Another advantage is that the second-order effect of getting past the person in the aisle is eliminated.  One drawback of this approach is that many people travel in groups and would likely not adhere strictly to this particular boarding policy.  In general, however, this approach did very well (as also found by \cite{vandenbriel} though for different reasons).

The modified optimal approach performed slightly better than, though almost identically to, the windows-middle-aisle approach.  This method does not depend as strongly on the length of the airplane.  So, for shorter airplanes it compares better while on longer planes it compares less well.  An advantage that the modified optimal approach has over boarding window-middle-aisle is that it allows passengers who are travelling together and sitting side-by-side to board at the same time without boarding out of order.

Random boarding, where the passengers positions in line are completely uncorrelated with their seat assignment, shares the advantage of the windows-middle-aisle method of spreading passengers throughout the length of the airplane.  Moreover, it is not disadvantaged, in implementation at least, by traveling groups.  The random method performed the similar to, but slightly worst than, the windows-middle-aisle method and the modified optimal method.  This demonstrates that the optimal approach, even with a significant fraction of people out of order can do at least as well as or better than the most efficient of the methods that employ boarding groups.

\begin{figure}
\begin{center}
\includegraphics[width=\fsize]{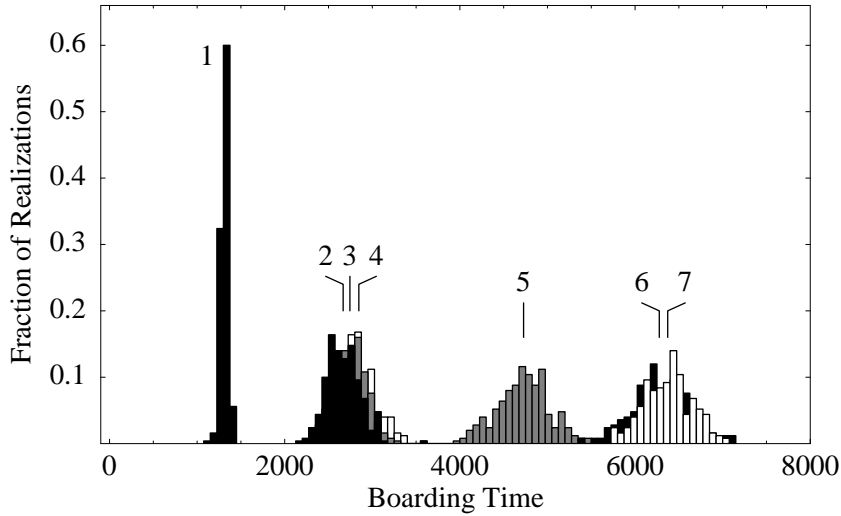}
\caption{Histogram of the loading times for 100 realizations of seven different boarding schemes.  The luggage loading time for each realization is drawn from a uniform distribution with a mean of 50 counts.  These are: 1) optimal boarding (mean boarding time: 1312 counts), 2) the modified optimal approach (2670), 3) the window-middle-aisle method (2750), 4) random boarding (2846), 5) ordered blocks from back to front (4727), 6) back to front with all passengers in order (6276), and 7) the worst-case---front to back with all passengers in order (6373).}
\label{blockresults}
\end{center}
\end{figure}

Figure \ref{blockresults} shows a histogram of the loading times for 100 realizations of seven different boarding schemes, each realization having different selections for the time it takes to load the luggage and different passenger ordering where applicable.  These include: 1) optimal boarding, 2) the modified optimal approach, 3) the window-middle-aisle method, 4) random boarding, 5) ordered blocks from back to front, 6) back to front with all passengers in order, and 7) the worst-case (front to back with all passengers in order).  We see from this figure that optimal boarding has, by far, the best improvment in loading times, nearly a factor of five faster than the worst case, more than a factor of three better than the ordered blocks, and it is more than a factor of two faster than the modified optimal, random, and windows-middle-aisle methods.  This improvement grows with the length of the airplane such that an airplane that seats 240 passengers (40 rows) will board over seven times faster than the worst case!

\section{Conclusion\label{conclusion}}

The results of this study, based upon the assumption that a passenger loading his luggage consumes the bulk of the time that it takes for him to be seated, identify the primary cause for delay in the boarding process as well as the best means to overcome these delays.  By boading passengers in a manner that allows several passengers to load their luggage simultaneously the boarding time can be dramatically reduced.  This result contradicts conventional wisdom and practice that loads passengers from the back of the airplane to the front.  Indeed, it shows that loading from the back to the front is hardly better than the worst case scenario.  The goal of an optimized boarding strategy should focus on spreading the passengers who are loading their luggage throughout the length of the airplane instead of concentrating them in a particular portion of the cabin.

By boarding in groups where passengers whose seats are separated by a particular number of rows, by boarding from the windows to the aisle, or by allowing passengers to board in random order one can reduce the time to board by better than half of the worst case and by a significant amount over conventional back-to-front blocks---which, while better than the worst case performed worse than all other block loading schemes.  The primary drawbacks for any of these methods is likely to be psychological instead of practical.  Groups of passengers who wish to board together would be an issue to investigate from both a customer satisfaction point of view and as a component in a more detailed model.

If a workable method to have passengers line up in an assigned order could be found---and it likely may be employed already, then there is the potential for a substantial savings in time.  Such a savings would most likely benefit flights between nearby cities where a particular airplane would make several trips in a given day since it might allow one or two additional flights.  Or, it might allow an airline to reduce the number of gates that it requires to meet its obligations since each gate would be cleared more rapidly.

While the generic features of this model are well understood, a real application of it would require some data so that it can be properly calibrated.  In particular, the distributions of luggage loading times, the fraction of people traveling in groups, the queueing habits of passengers, and empirical measurements of ``personal space'' are all pieces of information that are necessary to state these results in terms of actual times and distances instead of arbitrary time steps and lengths that are used by a computer model.

Regardless of the ultimate application of this technique, the establishement of a firm lower bound can be used to inform a decision maker of the worth of further improvements to a particular boarding strategy.  If an improvement could provide only a marginal gain while costing significant amounts money and time to implement then it is not likely to be worth the investment.  On the other hand, if a particular strategy is clearly failing to meet the demands of competition and customer satisfaction, then knowing just how much room there is for improvement could expedite changes.

In the end, the time that it takes to load passengers into the airplane affect not only the airline company and the airport, it also affects the passengers.  Few enjoy standing in line longer than necessary and fewer still enjoy sitting in an airplane longer than needed.  Faster boarding would be a significant improvement for all involved parties.

{\bf Acknowledgements:} J. Steffen acknowledges the generous support of the Brinson Foundation.  Fermilab is supported by the U.S. Department of Energy under contract No. DE-AC02-07CH11359.  A record of invention of this software and its results are on file at Fermilab.





\end{document}